\documentclass[12pt]{article}
\usepackage{graphicx}
\usepackage{rotating}
\usepackage{cite}
\usepackage{color}
\usepackage{amssymb}
\newcommand{\bm}[1]{\mbox{\boldmath $#1$}}
\newcommand{\babar}{{\it BABAR}}
\newcommand{\lijntje}[3]{(\begin{picture}(20,0)(4,0)
\end{picture})}
\begin{document}
\title{Comment on\\ ``Study of $D_{sJ}$ decays to $D^{\ast}K$\\ in inclusive
$e^{+}e^{-}$ interactions''}
\author{
Eef~van~Beveren$^{\; 1}$ and George~Rupp$^{\; 2}$\\ [10pt]
$^{1}${\small\it Centro de F\'{\i}sica Computacional,
Departamento de F\'{\i}sica,}\\
{\small\it Universidade de Coimbra, P-3004-516 Coimbra, Portugal}\\
{\small\it eef@teor.fis.uc.pt}\\ [10pt]
$^{2}${\small\it Centro de F\'{\i}sica das Interac\c{c}\~{o}es Fundamentais,}\\
{\small\it Instituto Superior T\'{e}cnico, Technical University of Lisbon,}\\
{\small\it Edif\'{\i}cio Ci\^{e}ncia, P-1049-001 Lisboa Codex, Portugal}\\
{\small\it george@ist.utl.pt}\\ [.3cm]
{\small PACS number(s): 14.40.Lb, 14.40.Ev, 13.25.Ft, 12.39.Pn}
}
\maketitle

\begin{abstract}
We comment on the recent observation
of the decay mode $D_{sJ}^{\ast}(2860)^{+}\to D^{\ast}K$ by
the {\it BABAR} \/Collaboration [B.~Aubert {\it et al.} ({\it BABAR}
\/Collaboration), arXiv:0908.0806],
and contest their peremptory conclusion that the data
exclude a $0^{+}$ assignment for the $D_{sJ}^{\ast}(2860)^{+}$.
In particular, we argue that the observed branching fraction
${\mathcal B}(D_{sJ}^{\ast}(2860)^{+}\to D^{\ast}K)/{\mathcal B}
(D_{sJ}^{\ast}(2860)^{+}\to DK)=1.1\pm0.15\pm0.19$
supports the existence of two largely overlapping resonances
at about 2.86~GeV,
namely a pair of radially excited tensor ($2^{+}$) and scalar ($0^{+}$)
$c\bar{s}$ states.
This scenario is further justified by comparing with the
corresponding excited charmonium states. Also other aspects of the
charm-strange spectrum are discussed.
\end{abstract}

In a very recent study of decays of charm-strange mesons \cite{ARXIV09080806},
the \babar\ \/Collaboration for the first time observed the decay
$D_{sJ}^{\ast}(2860)^{+}\to D^{\ast}K$.
When the $D_{sJ}^{\ast}(2860)^{+}$ meson was discovered by \babar\
\cite{PRL97p222001}, only the $DK$ decay mode was detected, which made an
assignment as a radially excited scalar $c\bar{s}$ meson plausible
\cite{PRL97p202001,PLB647p159,EPJC50p617}, though other configurations such
as a $3^-$ state \cite{PLB642p48,EPJC50p617,NPPS186p363} could not be excluded.
For a discussion of additional options, see Ref.~\cite{ARXIV09042453}.
The now observed $D^{\ast}K$ mode seems to exlude the $0^{+}$ scenario for the
$D_{sJ}^{\ast}(2860)^{+}$, as concluded by \babar.
However, in the following we shall show that the true situation
may be subtler, involving two overlapping resonances,
one scalar ($J^{P}=0^{+}$)
and one tensor ($J^{P}=2^{+}$) charm-strange meson.

The \babar\ angular analysis of the $D_{sJ}^{\ast}(2860)\to D^{\ast}K$
decays supports natural parity for the resonance,
i.e., $J^P=0^{+},1^-,2^{+},3^-,\ldots$.
Ruling out for the moment the scalar hypothesis,
experiment also seems to eliminate the solution
of Refs.~\cite{PLB642p48,NPPS186p363}, viz.\ $3^-$ ($1\,{}^{3\!}D_{3}$).
Namely, the measured \cite{ARXIV09080806} branching ratio
${\mathcal B}(D_{sJ}^{\ast}(2860)^{+}\to D^{\ast}K)/
{\mathcal B}(D_{sJ}^{\ast}(2860)^{+}\to DK)=1.1\pm0.15\pm0.19$
is not compatible with the value 0.39 predicted for a $3^-$ state
in Ref.~\cite{PLB642p48,NPPS186p363}.
{\it A fortiori}, a $1\,{}^{3\!}D_1$ assignment can be excluded, too,
as it would imply a branching ratio of only 0.06 in the latter model analysis.
As for the other $1^-$ state ($2\,{}^{3\!}S_{1}$), the predicted
\cite{PLB642p48,NPPS186p363} branching ratio of 1.23 {\it is} \/in agreement
with experiment (though see below).
However, one expects this vector meson to have a considerably lower mass,
close to 2.7~GeV, which makes the $D_{s1}^{\ast}(2700)^{+}$ \cite{PLB667p1},
now confirmed by \babar\ and denoted $D_{s1}^{\ast}(2710)^{+}$
\cite{ARXIV09080806}, a much better candidate.

These results appear to suggest that the $2^{+}$ ($2\,{}^{3\!}P_{2}$)
assignment for the $D_{sJ}^{\ast}(2860)^{+}$ is the most likely one.
However, the inevitable mixing of
the spectroscopic $2\,{}^{3\!}P_{2}$ and $1\,{}^{3\!}F_2$ states,
resulting in {\em two} \/$2^{+}$ states, complicates the analysis
of the tensor hypothesis.
Similar complications arise in the vector case,
with $2\,{}^{3\!}S_{1}$ and $1\,{}^{3\!}D_1$ mixing,
and the axial-vector case,
with $n\,{}^{3\!}P_1$ and $n\,{}^{1\!}P_1$ mixing, for any $n$,
when dealing with $K_1,D_1,D_{s1},\ldots$ mesons,
which are not $C$-parity eigenstates.
The latter mixing can be large \cite{PRD43p1679,EPJC32p493}.
Moreover, in Ref.~\cite{EPJC32p493}
it has been shown, that coupled channels naturally induce
such an axial-vector mixing, and tend to give rise to one state that
decouples to a large extent, while the other one is subject to a sizable
coupled-channel mass shift, of the order of 80~MeV downwards.
Since in our multichannel analysis of the $D_{sJ}^{\ast}(2860)^{+}$
\cite{PRL97p202001} we found a very similar mass shift
for the bare $2\,{}^{3\!}P_{0}$ state, we expect that a full
coupled-channel calculation of the, in our model,
degenerate bare $2\,{}^{3\!}P_{2}$ and $1\,{}^{3\!}F_2$ states
will result in one physical $2^{+}$ $c\bar{s}$ resonance
very close to the shifted $2\,{}^{3\!}P_{0}$ state,
and so to the observed $D_{sJ}^{\ast}(2860)^{+}$,
and another one at roughly 2.92~GeV.

Rather than relying upon the
coupling constants used in Refs.~\cite{PLB642p48,NPPS186p363},
since no information is available about the couplings of radially excited
heavy-light mesons to low-lying states \cite{PLB642p48}, we shall use the
formalism based on harmonic-oscillator expansions developed in
Ref.~\cite{ZPC17p135}. In this framework, all couplings of arbitrary
excited states to all two-meson decay channels can be straightforwardly
computed. Then, we for get the couplings squared
$\left( 2\,{}^{3\!}P_{2}\to D^{\ast}K\right)$ :
$\left( 2\,{}^{3\!}P_{2}\to DK\right)$ :
$\left( 1\,{}^{3\!}F_{2}\to D^{\ast}K\right)$ :
$\left( 1\,{}^{3\!}F_{2}\to DK\right)=$
21 : 14 : 4 : 6.
If we neglect phase-space effects for a first rough estimate,
we get the branching ratios
${\mathcal B}(\mbox{$2\,{}^{3\!}P_{2}$}\to
D^{\ast}K)/{\mathcal B}(\mbox{$2\,{}^{3\!}P_{2}$}\to DK)\sim 1.5$,
which is somewhat too large,
and ${\mathcal B}(\mbox{$1\,{}^{3\!}F_{2}$}\to
D^{\ast}K)/{\mathcal B}(\mbox{$1\,{}^{3\!}F_{2}$}\to DK)\sim 0.67$,
which is somewhat too small.
However, the situation changes dramatically for admixtures
because of interference.
When we assume an admixture of only 2\% $1\,{}^{3\!}F_2$
and 98\% $2\,{}^{3\!}P_{2}$, we find
${\mathcal B}(\mbox{$2\,{}^{3\!}P_{2}$}\to
D^{\ast}K)/{\mathcal B}(\mbox{$2\,{}^{3\!}P_{2}$}\to DK)\sim 2.1$,
whereas an admixture of 10\%, respectively 90\%, gives
${\mathcal B}(\mbox{$2\,{}^{3\!}P_{2}$}\to
D^{\ast}K)/{\mathcal B}(\mbox{$2\,{}^{3\!}P_{2}$}\to DK)\sim 3.2$.

Of course, phase space may change those numbers considerably,
by reducing the $D^{\ast}K$ rate, according to the usual
formulae based on perturbation theory.
However, also the predicted $DK$ rate may be considerably smaller,
since channels that are wide open tend to fade out
quite fast for further increasing momenta,
as confirmed by open-charm decays of charmonium \cite{ARXIV09080242}.
So if we take the branching ratios of 2--3 at face value,
there also is a problem with the $2^{+}$ assignment.
A possible way out is by supposing that there is
a scalar ($2\,{}^{3\!}P_0$) $c\bar{s}$ resonance as well,
largely overlapping with the $D_{sJ}^*(2860)^{+}$.
Since the scalar does not decay into $D^{\ast}K$,
a combination of tensor and scalar will have a smaller
branching ratio $D^{\ast}K/DK$.
In view of the above quoted result for this branching ratio
of the \babar\ \/Collaboration in Ref.~\cite{ARXIV09080806},
we thus need a large $2\,{}^{3\!}P_{0}$ contribution
to the $D_{sJ}^{\ast}(2860)^{+}$ resonance,
already for small to modest contributions of
an unavoidable $c\bar{s}$ $1\,{}^{3\!}F_2$ component to
the $c\bar{s}$ $2^{+}$ state at 2.860 GeV.

We can do an analogous analysis
for the $2\,{}^{3\!}S_{1}$ state.
Using the same scheme \cite{ZPC17p135} as above,
we get for the couplings squared
$\left( 2\,{}^{3\!}S_{1}\to D^{\ast}K\right)$ :
$\left( 2\,{}^{3\!}S_{1}\to DK\right)=$ 2 : 1,
resulting in a branching ratio of 2.
So in this scenario we would also need a $2\,{}^{3\!}P_{0}$
contribution in order to agree with experiment.
But more importantly, the $2\,{}^{3\!}S_{1}$ state is
unlikely to overlap with the $2\,{}^{3\!}P_{0}$,
as it should be significantly lighter.

Coming back to the $2^{+}/0^{+}$ scenario,
how reasonable is it to assume that
the $2\,{}^{3\!}P_{2}$ and $2\,{}^{3\!}P_{0}$ states overlap?
From the point of view of possible spin-orbit forces,
these are generally accepted to be small for radially
excited states and less important than e.g.\ coupled-channel effects.
On the other hand, browsing the PDG tables \cite{PLB667p1}
we do not find any example of a clearly established pair
of $2\,{}^{3\!}P_{2}$ and $2\,{}^{3\!}P_{0}$ states above
threshold for OZI-allowed decay.
However, charmonium may be an exception,
comprising the $\chi_{c2}(2P)$ with a mass of $3929\pm5\pm2$~MeV
and a width of $29\pm10\pm2$~MeV, and the $X(3945)$
with a mass of $3916\pm6$~MeV and a width of $40^{+18}_{-13}$~MeV
\cite{PLB667p1}.
Although most quantum numbers of the latter resonance
are undetermined so far, its positive $C$-parity \cite{PLB667p1}
makes it a very good candidate for the $2\,{}^{3\!}P_{0}$ $c\bar{c}$ state
(also see Ref.~\cite{PRD74p037501}).
So we may have here a pair of $2\,{}^{3\!}P_{2}$ and $2\,{}^{3\!}P_{0}$
resonances with central masses (marginally) within
each other's error bars.
Moreover, there may even be an indication of
two different $D_{sJ}^*(2860)^{+}$ resonances
in the very \babar\ data \cite{ARXIV09080806},
as their $DK$ fit tends to peak
at a slightly lower mass (2860~MeV) than the
$DK+DK^*$ (2860--2866~MeV) and $DK^*$ ones (2865~MeV)
(see Table~II of Ref.~\cite{ARXIV09080806}).

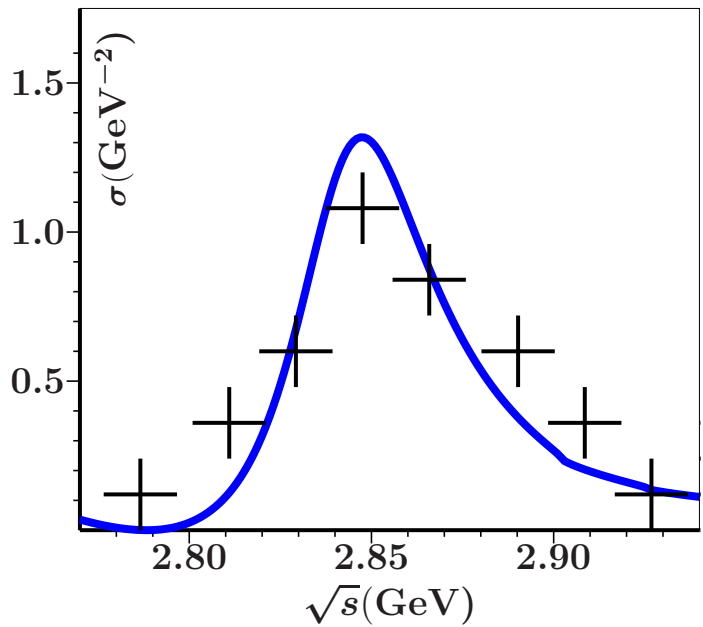
\begin{figure}[htbp]
\begin{center}
\begin{picture}(250,230)(0,-30)
\put(41.35,-5.52){\makebox(0,0)[tc]{\bf\large 2.80}}
\put(110.28,-5.52){\makebox(0,0)[tc]{\bf\large 2.85}}
\put(179.20,-5.52){\makebox(0,0)[tc]{\bf\large 2.90}}
\put(-5.52,56.43){\makebox(0,0)[rc]{\bf\large 0.5}}
\put(-5.52,112.86){\makebox(0,0)[rc]{\bf\large 1.0}}
\put(-5.52,169.29){\makebox(0,0)[rc]{\bf\large 1.5}}
\put(117.17,-20.07){\makebox(0,0)[tc]{\large\bm{\sqrt{s}}(\bf GeV)}}
\put(5.52,189.48){\makebox(0,0)[tl]{\begin{sideways}
\bm{\sigma}(\large\bf GeV\bm{^{-2}})\end{sideways}}}
\end{picture}
\end{center}
\caption[]{\small The experimental line shape
of $DK$ in the invariant--mass interval 2.77 -- 2.94 GeV
from data published by the BABAR Collaboration
in Ref.~\cite{PRL97p222001}
(in size adjusted to the theoretical curve)
and the result \lijntje{0}{0}{1}
of our coupled-channel model for inelastic $DK$ scattering.
}
\label{DK}
\end{figure}

In the coupled-channel calculation for $J^{P}=0^{+}$ of
Ref.~\cite{PRL97p202001} we definitely obtain a resonance
at 2.847 GeV with a width of about 50 MeV
(see Fig.~\ref{DK}).
A calculation for $J^{P}=2^{+}$ has not been performed yet.
But, since the bare states of $2^{+}$ and $0^{+}$ are degenerate
in our coupled-channel model and come at 2.925 GeV,
we expect to find a resonance for $2^{+}$
very near the $0^{+}$ resonance and with a comparable width.

In conclusion, we would like to stress the importance of the experimental
effort to further complete the $c\bar{s}$ spectrum, which has been receiving
an enormous boost by \babar\ in recent years.
However, it is crucial that the $D_{sJ}^{\ast}(2860)^{+}$
and the newly discovered $D_{sJ}(3040)^{+}$
get confirmed and their quantum numbers pinned down
by other collaborations.
In particular, a possible experimental disentangling
of the $D_{sJ}^{\ast}(2860)^{+}$ structure in a scalar and a tensor
part as suggested here
would be an enormous step forward in meson spectroscopy.

We are indebted to Prof.~Xiang Liu of Lanzhou University for drawing
our attention to the new \babar\ \/data. This work was supported by the
{\it Funda\c{c}\~{a}o para a
Ci\^{e}ncia e a Tecnologia} \/of the {\it Minist\'{e}rio da Ci\^{e}ncia,
Tecnologia e Ensino Superior} \/of Portugal, under contract
CERN/FP/83502/2008.

\newcommand{\pubprt}[4]{{#1 {\bf #2}, #3 (#4)}}
\newcommand{\ertbid}[4]{[Erratum-ibid.~{#1 {\bf #2}, #3 (#4)}]}
\def\EPJC{Eur.\ Phys.\ J.\ C}
\def\NPPS{Nucl.\ Phys.\ Proc.\ Suppl.}
\def\PLB{Phys.\ Lett.\ B}
\def\PRD{Phys.\ Rev.\ D}
\def\PRL{Phys.\ Rev.\ Lett.}
\def\ZPC{Z.\ Phys.\ C}

\end{document}